\documentclass[twoside,english,aps,prb,twocolumn,superscriptaddress]{revtex4-1}
\usepackage[T1]{fontenc}
\usepackage[utf8]{inputenc}
\usepackage{color}
\usepackage{textcomp}
\usepackage{amsmath}
\usepackage{amssymb}
\usepackage{graphicx}

\makeatletter
 
 \@ifundefined{textcolor}{}
 {%
   \definecolor{BLACK}{gray}{0}
   \definecolor{WHITE}{gray}{1}
   \definecolor{RED}{rgb}{1,0,0}
   \definecolor{GREEN}{rgb}{0,1,0}
   \definecolor{BLUE}{rgb}{0,0,1}
   \definecolor{CYAN}{cmyk}{1,0,0,0}
   \definecolor{MAGENTA}{cmyk}{0,1,0,0}
   \definecolor{YELLOW}{cmyk}{0,0,1,0}
 }

\usepackage[amssymb]{SIunits}\@ifundefined{definecolor}
 {\usepackage{color}}{}

\newcommand{\LSMO}{La$_{0.96}$Sr$_{2.04}$Mn$_{2}$O$_{7}$}

\makeatother

\usepackage{babel}
\begin{document}

\title{Antiferromagnetic Domain Structure in Bilayer Manganite}

\author{M. Garc\'ia-Fern\'andez}
\email{mgfernandez@bnl.gov}
\author{S. B. Wilkins}
\email{swilkins@bnl.gov}
\affiliation{Condensed Matter Physics and Materials Science Department, Brookhaven
National Laboratory, Upton, New York, 11973 USA}

\author{Ming Lu}
\affiliation{Center for Functional Nanomaterials, Brookhaven National Laboratory,
Upton, New York, 11973 USA}

\author{Qing'an Li}
\author{K. E. Gray}
\author{H. Zheng}
\author{J. F. Mitchell}
\affiliation{Materials Sciences Division, Argonne National Laboratory, Argonne,
IL, 60439}
\author{Daniel Khomskii}
\affiliation{Physikalisches Institut, Universität zu Köln, Zülpicher Straße 77,
D-50937 Köln, Germany}

\date{\today}

\begin{abstract}

We report a novel soft x-ray nanodiffraction study of antiferromagnetic
domains in the strongly correlated bylayer manganite \LSMO. 
We find that the antiferromagnetic domains are quenched, forming a unique domain pattern
with each domain having an intrinsic memory of its spin direction,
and with associated domain walls running along crystallographic
directions. This can be explained by the presence of crystallographic or magnetic imperfections locked
in during the crystal growth process which pin the antiferromagnetic
domains. The antiferromagnetic domain pattern shows two distinct types
of domain. We observe,
in one type only, a periodic ripple in the manganese spin direction
with a period of approximately 4~\micro\meter. We propose that the
loss of inversion symmetry within a bilayer is responsible for this
ripple structure through a Dzyaloshinskii-Moriya-type interaction. 

\end{abstract}

\maketitle

\begin{figure}[t]
\begin{centering}
\includegraphics[width=0.9\columnwidth]{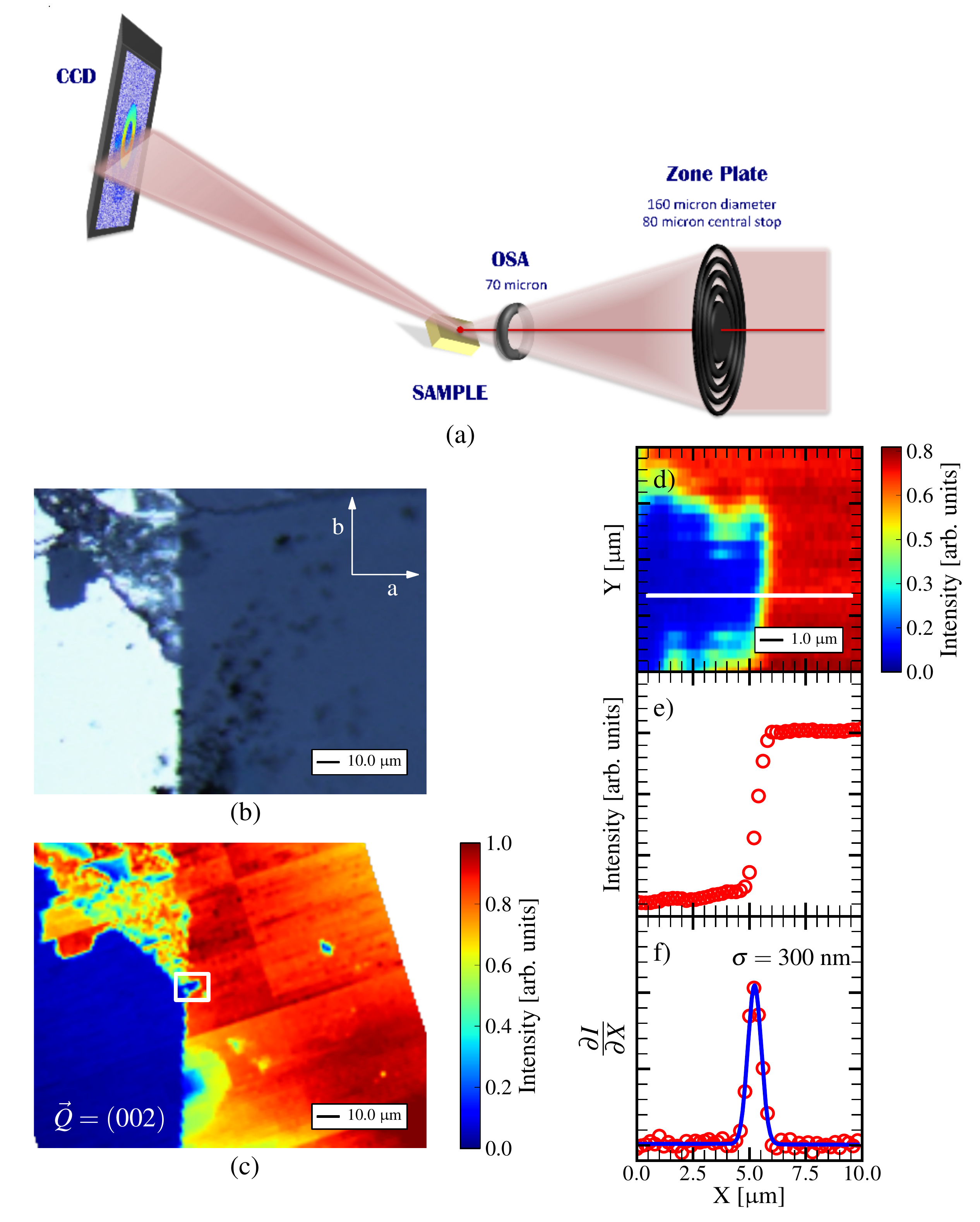} 
\par\end{centering}
\caption{
a) Schematic of the experimental setup. The soft x-ray beam is focussed
by a Fresnel zone plate. Scanning the latter, a real-space image can
be constructed. b) Image of the fiducial mark and sample obtained
using a polarized microscope. c) Spatial map of the intensity of the
(002) crystallographic reflection as measured in the same region of
the sample. This measurement is performed with a step size of 2 microns.
d) Same as the previous with a step size of 200 nm. A white line marks
the region of the map where the linear cut shown in Figure 1e is taken.
It should be noted that the linear variation of intensity in the maps
of the (002) reflection (Fig. 1b and Fig. 1d) are artifacts of the
scanning process, and reflect the scanning direction. f) Gaussian
fit to the derivative of the linear cut. The full-width-at-half-maximum
provides the upper limit to the beam size. This is 300 nm.}
\label{fig:002} 
\end{figure}

There are two main classes of magnetic ordering:
ferromagnetism and antiferromagnetism. Both types are exploited in
technological devices and find numerous applications. In both cases,
the local order forms microscopic domains. While ferromagnetic (FM) domain
formation is well understood\cite{Hubert1998,Neel1955,LandauLifshitz1984},
antiferromagnetic (AFM) domains remain mysterious mostly
due to a lack of techniques which can spatially resolve the AFM order.
This is unfortunate, since antiferromagnetism is an ubiquitous magnetic ordering important for
both fundamental research and practical applications. It is used
to pin ferromagnetic layers in exchange bias systems\cite{Scholl2000,Nogues2005}
and can cause spin-valve-like behavior in multilayers of an AFM layer
and a non-magnetic layer\cite{Park2011}. From a fundamental physics
perspective, it is also one of the most common ground states in condensed
matter physics, especially in transition metal oxides. A prominent
example being AFM stripe order in doped, superconducting cuprates\cite{Tranquada2008}.
Further, AFM domains play an important role in defining the operation
of any device. For example, in nanoscaled exchange bias systems they
determine the strength of the exchange bias \cite{Mauri1987,Malozemoff1987,Miltenyi2000,Nowak2002}.
Antiferromagnets, unlike ferromagnets, have no net demagnetizing field
and therefore the formation of domains is not required by energy considerations,
although one could argue that entropy suggests their
presence. Basic questions therefore remain to be answered, including
the size and structure of AFM domains and if they are quenched, or
if they can easily be annealed. 

The study of AFM domains is extremely challenging. The absence of
any net magnetic moment prohibits the use of most magnetic imaging
techniques. Recently, Photoemission Electron microscopy (PEEM) combined
with X-ray Magnetic Linear Dichroism (XMLD) has been used to image
AF domains \cite{Stoehr1999,Scholl2000}. Here we present the novel
technique of soft x-ray resonant nano diffraction, which has considerable
advantages over PEEM, and is able to spatially resolve long range
electronic ordering\cite{Wilkins2003,Abbamonte2004,Garcia-Fernandez2009},
with a resolution of better than 300~nm (See Fig.~\ref{fig:002}a.) and
a sampled depth of greater than $1000$~\AA. These results are therefore representative of the bulk\cite{Wilkins2003}.
Soft x-ray resonant nano-diffraction is uniquely powerful in that
it can simultaneously map, and separate, the spatial profile of the
structural, magnetic, orbital and electronic correlations associated
with a particular atomic species. Previous spatially resolved x-ray
measurements have either used hard x-rays, which are not directly
sensitive to the underlying electronic order\cite{Kim:202505} or have used
x-ray optics which cannot obtain the sub-micron resolution reported
here\cite{Hiraoka2011}. It is only the combination of soft x-ray scattering with
a sub-micron probe which allows for the study of phase segregation, inhomogeneities and domains
in 3\textit{d} transition metal oxides. 

Experiments we carried out at the X1A2 beamline of the NSLS
at Brookhaven National Laboratory. The soft x-ray beam was focussed
by a Fresnel zone plate with 160~\micro\meter\ diameter,
90~\nano\meter\ outermost zone width and a 80 ~\micro\meter\ central
stop. A 70~\micro\meter\ pinhole was used as an order
sorting aperture. Both devices were mounted on $x$,$y$,$z$ positioning stages.
The incident beam was scanned with respect to the sample by moving
both the zone-plate and the order sorting aperture in unison and
keeping the sample fixed at the Bragg condition. An in-vacuum charge
coupled device (CCD) detector was used to detect the scattered x-rays. To ensure that
the same region of sample was studied throughout
and to provide a convenient edge for focussing and beam-size determination,
fiducial marks were placed on the surface of the sample by a lift-off
process similar to Ref.~\onlinecite{Hatzakis1980}. Two chromium lines of
200~\micro\meter\ width and approximately 1000~\AA\ thickness
(sufficient to attenuate the soft x-ray beam) were deposited on the
surface of the sample. An optical microscopy image of the portion
of the sample studied is shown in Fig~\ref{fig:002}b. Here, the
Cr deposited on the surface of the sample appears in lighter color.
The Cr lines run parallel to the $\langle100\rangle$ and $\langle010\rangle$ crystallographic
directions. The azimuthal angles were defined as the angle between
the $[100]$ direction and the vertical projection of the incoming
x-ray.

Polycrystalline bilayer manganite La$_{0.96}$Sr$_{2.04}$Mn$_{2}$O$_{7}$
was prepared by conventional solid reaction: La$_{2}$O$_{3}$, SrCO$_{3}$,
and MnO$_{2}$ were mixed according to stoichiometry. The mixtures
were reacted in the air at 1000 \textdegree{}C, 1100 \textdegree{}C,
1200 \textdegree{}C, and 1300 \textdegree{}C for 24 hours, respectively,
with intermediate grinding. The resulting single-phase mixtures were
pressed into rods and sintered at $1400^\circ$C for 24~hours
in air. Single crystals of La$_{0.96}$Sr$_{2.04}$Mn$_{2}$O$_{7}$
were grown using floating zone (FZ) method in an IR image furnace
under a flowing 20\% oxygen ambient. \LSMO\ undergoes a transition from
a paramagnetic to an A-type AFM at T$_{N}$ = 205 K, in which ferromagnetic
Mn-O layers, with the spins lying in the plane, are antiferromagnetically
coupled along the \textit{c}-axis, with the two layers of a bilayer
also antiparallel. This gives rise to a (001) superlattice reflection
at a location in reciprocal space which is forbidden for fundamental
Bragg scattering. Exploiting this reciprocal space selectivity, we
can image both the crystal structure and the AFM domain structure
simply by tuning to the (002) and (001) reflections respectively. 

The x-ray resonant nano-diffraction measurements were performed as
follows; a single CCD image was collected at each point in the two
dimensional map,
and the integrated intensity was extracted. The step-size used for
each map of intensity is adopted as our resolution since it is much
larger than the x-ray beam size. To experimentally
determine an upper bound on the ultimate resolution of our setup,
a smaller map was measured with a step size of 200~\nano\meter\ 
(Fig~\ref{fig:002}d). By extracting a line cut from this map (Fig.~\ref{fig:002}e)
and taking the differential, (Fig.~\ref{fig:002}f) a profile of
the focussed x-ray beam, convolved with the chromium feature was obtained.
Fitting this result to a Gaussian function we obtain an upper bound
of $\sigma$=300~\nano\meter.  

To measure the spatial image of the crystal lattice,  the (002) reflection was mapped at an energy of 770~eV. The results
are shown in Fig. 1c and Fig. 1d. These maps show that the sample is homogeneous to the resolution of
our measurement over length scales of 10's of \micro\meter\ and rules
out the possibility of ingrowths that are often present 
in such 327-systems.

\begin{figure}[t]
\begin{centering}
\includegraphics[width=0.9\columnwidth]{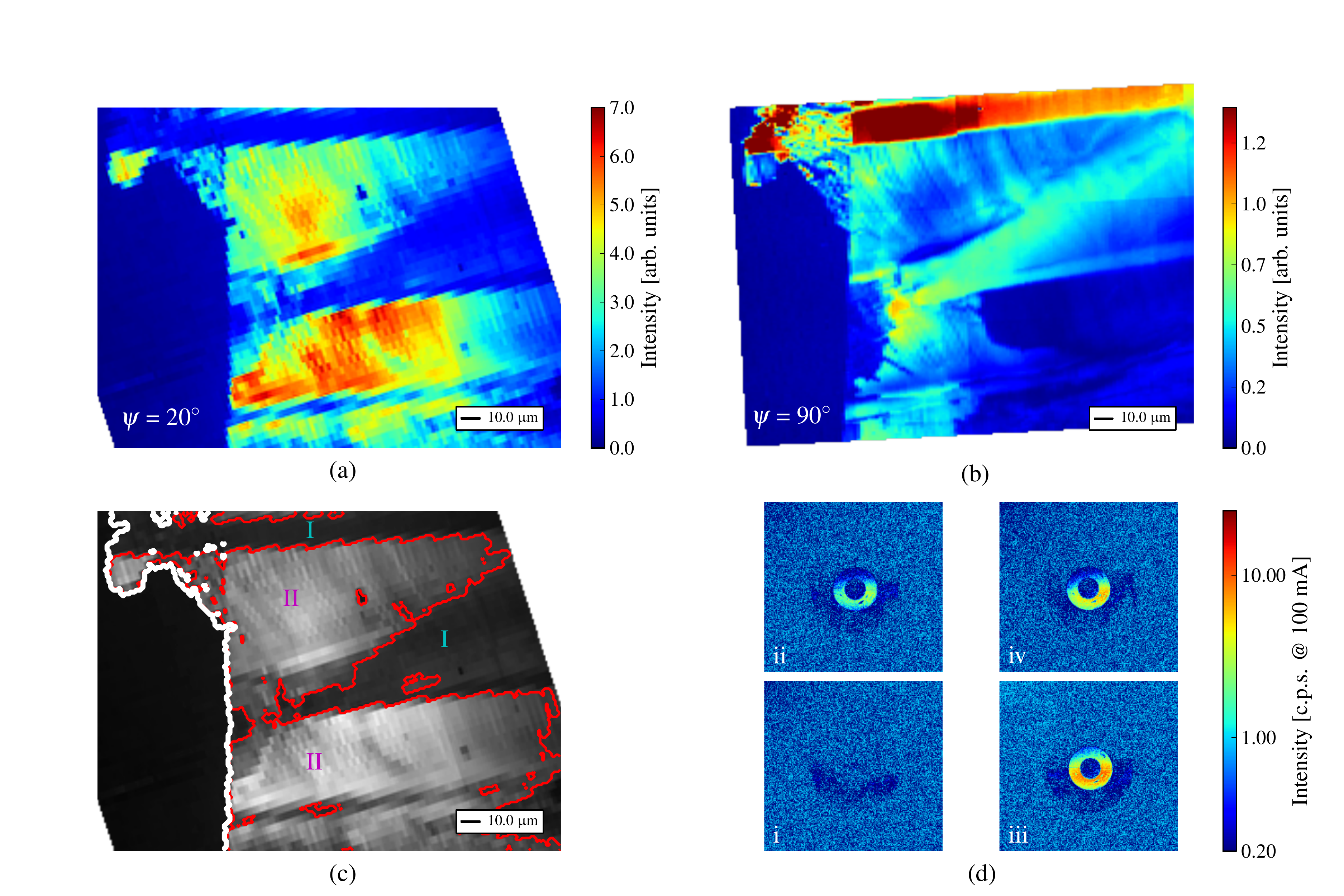} 
\par\end{centering}

\caption{a) Map of the antiferromagnetic (001) reflection
in the vicinity of the fiducial mark with a beam size of less than
300 nm and a step size of 2 microns for the azimuthal angle $\psi=20^{o}$.
Figure 2b shows the same region for the azimuthal angle $\psi=90$$^{o}$.
These data are collected with a step size of 1 micron. Figure 2c shows
a schematic of the different regions observed in the mapping of the
(001) reflection. Figures di, dii, diii and div show individual CCD
images of the (001) reflection through the zone plate for the different
regions in the AFM map in Figure 2b.}

\label{fig:001}
\end{figure}

The spatial dependence of the A-type AFM was then measured by tuning
the incident x-rays to the Mn $L_{3}$-edge (642 eV) \cite{Wilkins2003}
and measuring the magnetic (001) reflection. The x-ray cross-section is dependent on the direction of the local
moment and the incoming (and outgoing) x-ray polarization. In this
experiment, the incident x-ray polarization was horizontal, perpendicular
to the scattering plane ($\sigma$) which results in the maps being
sensitive to the projection of the spins along the scattered x-ray
beam \cite{Hill1996}.

Figure~\ref{fig:001}a shows the spatial dependence of the AFM (001)
reflection at T = 85 K. In contrast to the (002) crystallographic
reflection (Fig.~\ref{fig:002}b), we observe that the A-type AFM
is not homogeneous. Rather, two distinct regions are observed, labelled
type I and II and denoted in red in Fig. 1c. Domain walls run along
the $[100]$ and the $[110]$ directions, but not along $[010]$.
There are three main characteristics of these regions. Firstly, the
two regions are not equally populated over the measured crystal volume,
with type II occupying approximately two thirds of the investigated
volume. Secondly, type II regions exhibit some intriguing fine structure
that is absent in type I. Finally, for this orientation of the sample,
the intensity of the A-type (001) reflection is much lower in type
I than in type II. There are two possible explanations for this. Either
the difference in intensity is due to a reduction of the A-type AFM
order parameter in type I regions, or there is a different orientation
of the ordered moment relative to the incident x-ray beam in the two
regions. In order to distinguish between these two scenarios we measured
the same area, after rotating the sample $\sim70^{\circ}$ around
the surface normal. If a diminished order parameter is assumed and
all areas have the same direction of the ordered moment, then only
a change in the global intensity would be observed. Fig.~\ref{fig:001}b
shows the same area after the rotation. An inversion of the intensity
ratio between regions of type I and II is clearly visible. This result
indicates that both regions are AFM domains and that the ordered moment
forms an angle of approximately 90$^{o}$ between regions I and II.
The Cr fiducial mark is parallel to the {[}100{]} crystallographic
direction. This implies that the ordered moment is parallel to the
{[}100{]} and {[}010{]} direction in regions of type I and II respectively.
(Note that there may also exist AFM domains which have the same ordered
moment direction but a different phase. Our method does not allow
us to discriminate these domains).

\begin{figure}[t]
\begin{centering}
\includegraphics[width=0.9\columnwidth]{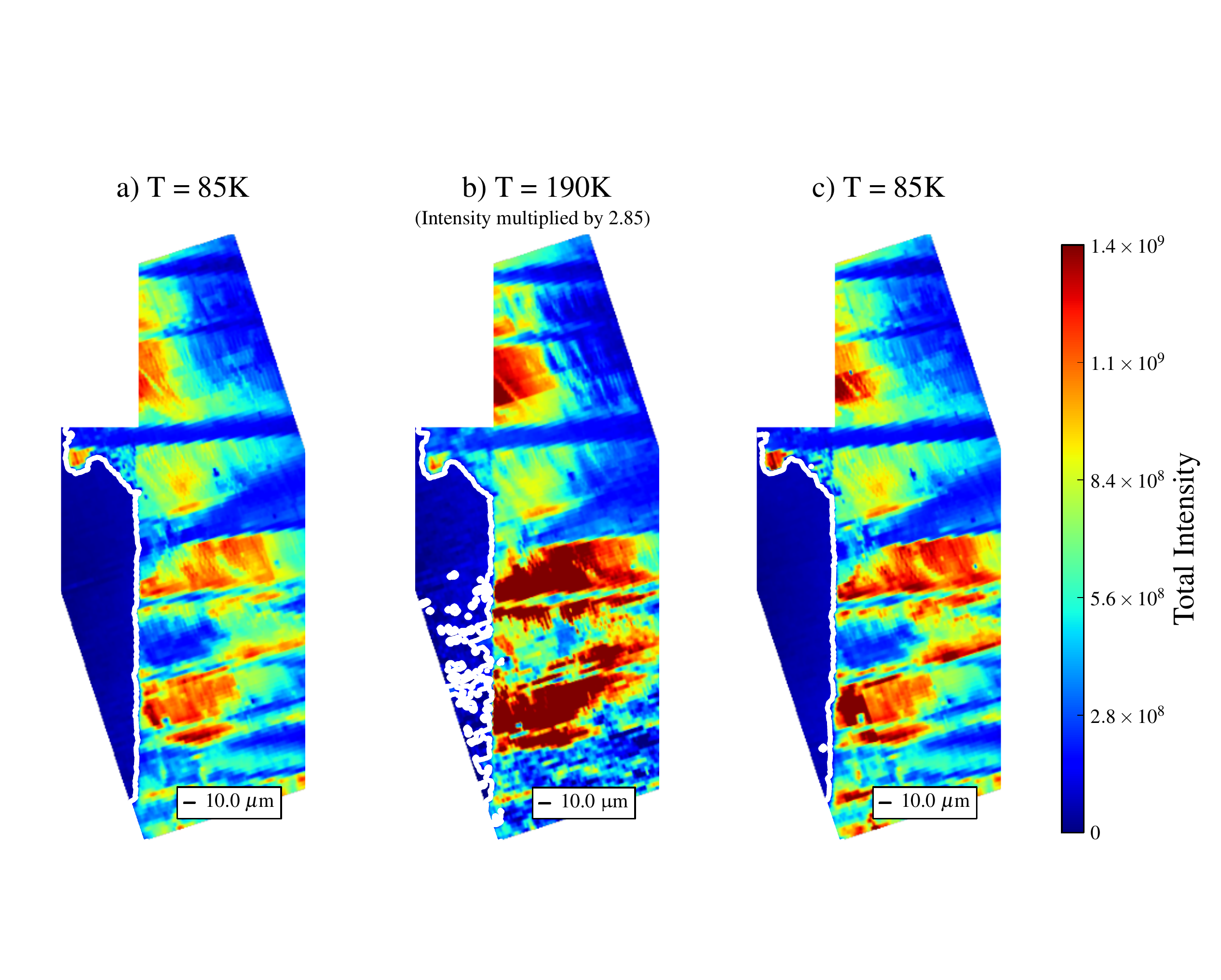} 
\par\end{centering}

\caption{a) (001) AFM map measured at
85 K. b) (001) AFM map measured at 190 K close to the Néel temperature.
c) (001) AFM map measured at 85 K after a full temperature cycle in
which sample was warmed up above the transition temperature, then
cooled down to 12 K and finally warmed up back to 85 K. The Cr fiducial
mark appears outlined in white for clarity. The intensity of the measurement
at 190 K was multiplied by 2.85 so that the image could be shown on
the same intensity range as the other two measurements. }
\label{fig:tdep} 
\end{figure}

We now turn to the evolution of the A-type AFM domains with temperature.
A map at 85 ~K taken over a large region and at the original orientation,
is shown in Fig.~\ref{fig:tdep}a. Fig.~\ref{fig:tdep}b shows a
map taken on warming to 190~K, close to the Néel temperature of 205~K.
Comparing Fig.~\ref{fig:tdep}a and Fig.~\ref{fig:tdep}b we find
that the overall intensity of each region has decreased to 36\% of
the value at 85 K, consistent with the reduction of the global order
parameter for these two temperatures \cite{Wilkins2003}. The actual
domain pattern however, once rescaled by the overall intensity, is
indistinguishable between the two cases. We further warmed the sample
to 300~K, at which point no signal was observed from any point on
the sample. To eliminate possible hysteresis, the sample was then
cooled to 12~K and warmed back to 85~K. The map obtained after this
thermal cycle is shown in Fig.~\ref{fig:tdep}c. The same global
intensity was recovered as at 85~K. Surprisingly, we find that the
domain pattern was not altered by this thermal cycling; the two measurements
are indistinguishable. This demonstrates that both the A-type AFM
domain patterns and the spin directions are quenched. We conclude
that the overall domain pattern is governed by features intrinsic
to the particular crystal and that are not annealed by warming to
300~K. Strain or internal fields could be involved in the appearance
of the observed domain pattern, which may not be visible in the measurement
of the (002).

\begin{figure}[t]
\begin{centering}
\includegraphics[width=0.9\columnwidth]{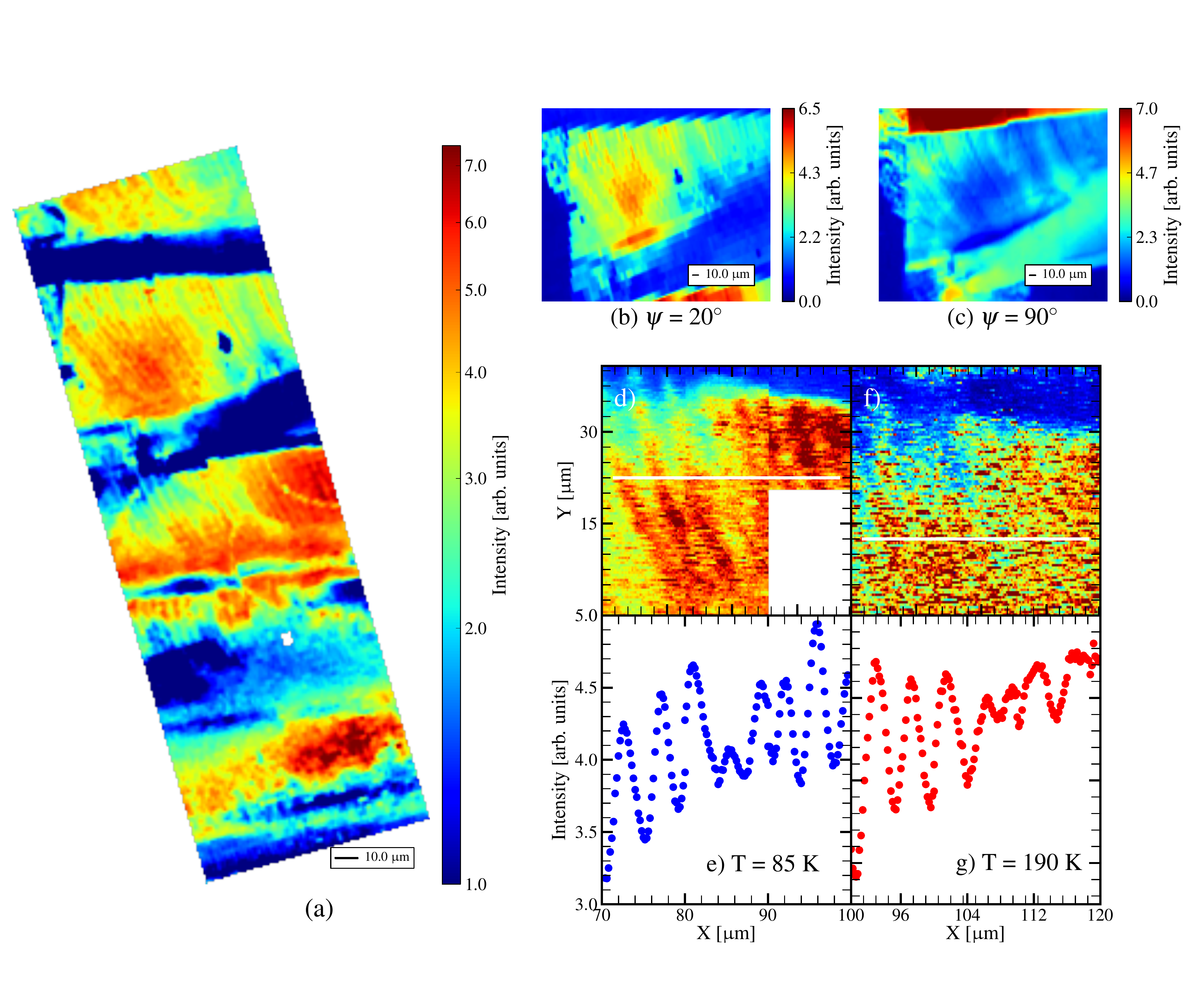} 
\par\end{centering}

\caption{a) Map of
the antiferromagnetic (001) reflection measured with a beam size of
less than 300 nm and with a step size of 1 micron. Two different types
of domain are observed. 4b) Zoom of the ripples observed in region
II measured with the same conditions as the overview for an azimuthal
angle of $\psi=90^{o}$. 4c) Same as b) with $\psi=20^{o}$. 4d) Zoom
of the ripples observed in domains of type II measured with a step
size of 200 nm and a temperature of 85 K. 4e) Linear cut of these
ripples marked with a white line in 4d. 4f) and g) Same as d) and
e) but at T=190 K.}

\label{fig:ripples} 
\end{figure}

We address now the fine structure observed within the type II domains
(Fig. 4a). In order to further characterize these {}``ripples'',
a map with much higher resolution (step size 200~\nano\meter) was
measured from a small area of type II (Fig. 4d), Figure~\ref{fig:ripples}e
shows a periodic modulation with a period of $\sim4$~\micro\meter\ and
an intensity modulation of $\sim40$\%. A similar analysis at T =
190~K, showed that while the overall intensity has decreased, consistent
with the reduced value of the order parameter at this temperature,
the ripples are still observed. From the line-out in Fig.~\ref{fig:ripples}g
the period and amplitude of the ripple modulation is almost identical
to that at 85~K. This modulation could be due to a spatial variation
of the order parameter (the magnitude of the ordered moment), or of
the ordered moment direction. Applying the same experimental method
used earlier, we compare images at two different rotations around
the surface normal. Comparing Fig.~\ref{fig:ripples}b and Fig.~\ref{fig:ripples}c
we observe that the part of the ripples which are of high intensity
at $\psi=20^{\circ}$ are of low intensity at $\psi=90^{\circ}$,
while regions of low intensity at $\psi=20^{\circ}$, have the same
intensity at $\psi=90^{\circ}$. Such behavior is only consistent
with the intensity modulation arising from the ordered moment direction
oscillating between the $[100]$ and $[110]$ directions.

One appealing explanation for this is that the long-period ``ripple''
structure is caused by the presence of the Dzyaloshinskii-Moriya (DM)
interaction arising from the loss of inversion symmetry at each bilayer.
It is well known that in multiferroics \cite{Khomskii2009} and even
at the metal-vacuum interface \cite{Bode2007,Ferriani2008} the loss
of inversion symmetry often results in the system forming a long period
cycloidal-type magnetic structure. In these materials, a cycloidal
magnetic structure can induce an electric polarization and \textit{vice
versa}. By analogy with both these cases, the loss of inversion symmetry
along the \textit{c}-axis at each bilayer could lead to a cycloidal
component to the magnetic structure, with moments rotating in the
b-c plane. The consequences of this within the context of our measurements
would be to observe an oscillating moment between the $[100]$ and
$[110]$ directions, since the x-ray measurements are only sensitive
to the component of the ordered moment along the exit wave vector.
The problem in this explanation is the large period of the ripples
($\sim$4 microns). The longest period known in multiferroic systems
is the periodic spin modulation in BiFeO$_{3}$ of approximately
630~\AA, which is almost 100 times shorter than the ripple period observed
here. A larger value of dielectric constant would be expected
in the present case (\LSMO\ is close to metallic) which would lead
to a sufficiently large increase of the period of the cycloid. This,
however, remains an open question.

The experimental data shown here present some interesting facts about
the AFM domain structure in \LSMO. We have observed the existence
of two different A-type AFM regions in the sample. These regions have
the same $\vec{Q}$-vector, but ordered moment directions that differ
by $90^{\circ}$. The domain walls separating these two regions run
along the $[100]$ and $[110]$ directions but not along $[010]$.
Further, the two regions have an unequal population with region II
occupying about 70\% of the measured volume. When studied in detail,
Regions I and II were found not to be symmetry related and therefore
are not AF domains in the true sense. For example, we found that type
II regions exhibit fine structure, absent in type I. These differences
between regions with the ordered moment along $[100]$ and $[010]$
are, \emph{a priori}, not to be expected from the known \LSMO $\;$crystal
structure. The observed behavior strongly suggests that the $a$-
and $b$-axes are not equivalent and that therefore the system is
orthorhombic and regions I and II are better classified as different
regions of A-type AFM ordering. We have observed that both the domain
pattern and spin direction are quenched and the material exhibits
a kind of `memory' which cannot be annealed upon warming to 300 K.

With such a quenched domain structure, we are left wondering what
determines the domain pattern? The measurements of the $(002)$ Bragg
reflection show that, at least on length scales of $\sim300$~\nano\meter,
the sample is crystallographically homogeneous, that is the intensity
of the reflection varies to less than 15\% and the position is unchanged
within errors. The observed domain pattern and behavior is different
to that observed in some other well studied antiferromagnets. In rhombohedral
NiO \cite{ROTH1960,SLACK1960}, while the type of domain walls that
separate regions of different magnetic $\vec{Q}$ vector, do indeed
run along crystallographic directions, the domains are far larger
(over several mm in size), and warming above $T_{N}$ is sufficient
to anneal the domain walls \cite{ROTH1960}. In addition, in NiO the
regions of different spin directions in a single magnetic $\vec{Q}$-vector
region form a nearly random pattern in zero applied magnetic field
\cite{ROTH1960}. Overall, \LSMO\ appears to behave rather differently.
One possible explanation is that the domain wall positions are governed
by the presence of crystallographic or magnetic imperfections locked
in during the crystal growth process.

We hope that the present results will stimulate further theoretical
and experimental work on the origin of this intriguing domain pattern
and the role that these previously inaccessible images of domain structures
play in the evolution of the phase diagrams of strongly correlated
electron systems, and even new emergent properties.

\begin{acknowledgments}
The authors would like to thank John Hill, Jing Tao and Simon Billinge
for stimulating discussions. Technical assistance from William J.
Leonhardt, D. Scott Coburn, Sue Wirick and William Schoenig is especially
noted. We would also like to acknowledge the support of Xiao Tong
with AFM measurements. This research was funded by the Department
of Energy, Office of Basic Energy Sciences, under Contract No. DE-AC02-98CH10886
at Brookhaven National Laboratory and DE-AC02-06CH11357 at Argonne
National Laboratory. Part of this research was carried out at the
Center for Functional Nanomaterials, Brookhaven National Laboratory.
\end{acknowledgments}

\bibliography{LSMO_Imaging}

\end{document}